\documentclass{article}[12pt]
\usepackage{epsf}
\usepackage[latin1]{inputenc}
\usepackage{amsmath,amsfonts,amssymb}

\def\beq{\begin{equation}}
\def\eeq{\end{equation}}
\def\bea{\begin{eqnarray}}
\def\eea{\end{eqnarray}}
\def\nn{\nonumber}

\def\neq{\not=}

\date{}

\begin{document}

\begin{titlepage}
\begin{center}
{\large\bf
Relating Jack Wavefunctions to $\textrm{WA}_{k-1}$ theories 
}\\[.3in] 

{\bf Benoit Estienne$^{1}$ and  Raoul\ Santachiara$^{2}$}\\
	$^1$ {\it LPTHE, CNRS, UPMC Univ Paris 06 \\
Bo\^ite 126, 4 place Jussieu, F-75252 Paris Cedex 05\\
            e-mail: {\tt estienne@lpthe.jussieu.fr}}\\

	$^2$  {\it LPTMS,CNRS,UMR 8626,
             Universit\'e Paris-Sud,\\ 
             B\^atiment 100\\
             91405 Orsay, France. \\
    e-mail: {\tt raoul.santachiara@lptms.u-psud.fr}. }\\
\end{center}
%\vskip .04in
\centerline{(Dated: \today)}
\vskip .2in
\centerline{\bf ABSTRACT}
\begin{quotation}
The $(k,r)$ admissible Jack polynomials, recently proposed as many-body wavefunctions for non-Abelian fractional quantum Hall systems,  have been  conjectured to be related to  some correlation functions of the minimal model $\textrm{WA}_{k-1}(k+1,k+r)$ of the $\textrm{WA}_{k-1}$ algebra.
By studying the degenerate representations of this conformal field theory, we provide a proof for this conjecture.

%\begin{quotation}
\end{quotation}
\end{titlepage}

\vskip 0.5cm
\noindent
%\pacs

\section{Introduction}
The conformal symmetry is extremely powerful in the study of two-dimensional (2D) massless quantum field theories because the algebra of its generators, the Virasoro algebra, is infinite dimensional \cite{diFrancesco,Dotsi_cours}. The Hilbert space of the simplest  family of conformal field theories (CFTs) is  built from the representations of this algebra. In these theories the correlation functions satisfy differential equations which are related to conformal invariance and to the degeneracy of the representations of the Virasoro algebra \cite{diFrancesco,Dotsi_cours}. In particular, among these representations, there are fields which obey   a so-called second order null-vector condition. This condition implies that any correlation function involving these fields satisfies a second order differential equation.  As it has been pointed out in different works (see for instance \cite{cardy1,cardy2} and references therein),  these differential equations   can be related to differential operators which define the Calogero-Sutherland  quantum Hamiltonian (\cite{Calogero}). The eigenstates of these many-body Hamiltonians, which describe $n$  particles interacting with a long range potential with  coupling $\alpha$,   are Jack polynomials (Jacks, defined below) \cite{Sutherland71,Sutherland72}. These are symmetric functions  in $n$ variables indexed by partitions $\lambda$ and by the parameter $\alpha$.

For a given number of variables $n$ and for each pair of positive integers $(k,r)$ such that $k+1$ and $r-1$ are coprime, one can define a Jack, which we denote $P^{(k,r)}_n$,  characterized  by a negative rational parameter $\alpha=-(k+1)/(r-1)$ and by some specific  partition (given below). The $P^{(k,r)}_{n}$ Jack  satisfies  the so called $(k,r)$ clustering conditions \cite{FJMM,FJMM2, BernevigHaldane1,BernevigHaldane2}, i.e. it  does not vanish when $k$ variables have the same value  but it vanishes with power $r$ when  the $k+1$-st variable approches a cluster of $k$ particles. Due to these properties, these Jacks have been considered as trial many-body wavefunctions  for  fractional quantum Hall ground states. In particular, the $P^{(k,r)}_n$ states  have been proposed  as possible generalizations of $\mathbb{Z}_k$ Read-Rezayi states \cite{MooreRead,ReadRezayi} for describing new non-Abelian states \cite{BernevigHaldane1,BernevigHaldane2}.  
 
In  \cite{FJMM,FJMM2} it has been conjectured that the $P_{n}^{(k,r)}$ can be written in terms of certain correlators of a family of  CFTs, the $\textrm{WA}_{k-1}$ theories.  This is a family of CFTs  with $W$ extended symmetry: in addition to the conformal symmetry, generated by the stress-energy tensor $T(z) = W^{(2)}(z)$ of spin $s=2$, the $\textrm{WA}_{k-1}$ theories enjoy additional symmetries generated by a set of chiral currents $W^{(s)}$ of spin $s=2,\dots,k$ \cite{Wtheory,Schoutens}. The $\textrm{WA}_{1}$ algebra  coincides with the Virasoro one.  The representations of the $\textrm{WA}_{k-1}$ algebras are naturally associated to the simple Lie algebra $A_{k-1}$ and the serie of minimal models $\textrm{WA}_{k-1}(p,q)$  is indexed by two integers $p$ and $q$ \cite{Wtheory,Schoutens}.  The theories $\textrm{WA}_{1}(p,q)$ correspond to the Virasoro minimal models $M(p,q)$.  For general $k>2$, however,   the $\textrm{WA}_{k-1}$ theories are much more complicated. This is  mainly because, contrary to the case of the Virasoro algebra,  the null-vector conditions characterizing a degenerate field do not in general lead to differential equations for the corresponding correlation functions. For these reasons, the problem of computing correlation functions of these higher spin symmetry CFTs is an hard problem \cite{TFTcf,TFTcf2}.

The conjecture that some correlation functions of the $\textrm{WA}_{k-1}$ theory can be written in terms of a single Jack polynomial is then quite remarkable. To be more precise, the conjecture states  that the $P^{(k,r)}_{n}$ Jack is directly related to the $n-$point correlation functions of certain fields (given below) of the theory $\textrm{WA}_{k-1}(k+1,k+r)$.  This is known to be true for the case $k=2$ corresponding to the Virasoro algebra. For general $k$, strong evidences supporting this conjecture have been provided in \cite{BernevigHaldane1,BernevigHaldane2,BernevigW,Ardonne} but a rigorous  proof was still missing.

We consider the $n-$point correlation function of certain operators of the $\textrm{WA}_{k-1}(k+1,k+r)$ theory. 
Using the approach described in \cite{TFTcf}, we show that these correlation functions satisfy a second-order differential equation which is directly related to the Calogero-Sutherland  quantum Hamiltonian. This provides a  proof for  the above conjecture.

\section{Symmetric polynomials and Jack Polynomials at $\alpha=-(k+1)/(r-1)$}
\label{Jacks}
A general characterization of symmetric polynomials which vanish when $k+1$ variables take the same value was initiated in the work of Feigin et al. \cite{FJMM}. In this section we briefly review their results and fix our notations.

The Jack polynomials $J^{\alpha}_{\lambda}(z_1,\cdots,z_n)$ are symmetric polynomials of $n$ variables depending rationally on a parameter $\alpha$ and indexed by partitions $\lambda$, $\lambda=[\lambda_1,\lambda_2\dots \lambda_n]$ where the $\lambda_i$ are a set of positive and decreasing integers $\lambda_1 \geq \lambda_{2} \geq \cdots \geq \lambda_n \geq 0$. For more details on Jack polynomials see \cite{Macdonald}. Defining the monomial functions $m_\lambda$ as:
\begin{equation}
m_{\lambda}(\{z_i \})=\mathcal{S} (\prod_i^n z_i^{\lambda_i}) 
\end{equation}
where the $\mathcal{S}$ stands for the symmetrization over the $n$ variables, the expansion of a Jack over the $m_{\lambda}$ basis takes the form \cite{Macdonald}:
\begin{equation}
J^{\alpha}_{\lambda}=m_{\lambda}+\sum_{\mu< \lambda} u_{\lambda \mu}(\alpha)m_{\mu}.  
\label{expan_jack}
\end{equation}
The dominance ordering $\mu\leq \lambda$ in the sum is defined as $\mu_1+\cdots+\mu_i\leq \lambda_1+\cdots\lambda_i$ ($1\leq i\leq n)$.
The Jacks $J^{\alpha}_{\lambda}$ are eigenfunctions of a Calogero-Sutherland Hamiltonian 
$\mathcal{H}^{\mbox{CS}}(\alpha)$ of coupling $\alpha$ \cite{Sutherland71,Sutherland72}:
\begin{equation}
\mathcal{H}^{\mbox{CS}}(\alpha)=\left[\sum_{i=1}^{n}(z_i\partial_i)^2+\frac{1}{\alpha}\sum_{i<j}\frac{z_i+z_j}{z_i-z_j}(z_i\partial_i - z_j \partial_j)\right]
\label{CS}
\end{equation}
More specifically, one has \cite{Sutherland71,Sutherland72}:
\begin{equation}
\mathcal{H}^{\mbox{CS}}(\alpha) J^{\alpha}_{\lambda}(z_1,\cdots,z_n)=\varepsilon_{\lambda}J^{\alpha}_{\lambda}(z_1,\cdots,z_n)
\label{CS2}
\end{equation}
where the eigenvalues $\varepsilon_{\lambda}$ are given by the following formula:
\begin{equation}
\varepsilon_{\lambda}=\sum_{i}^n \lambda_i\left[\lambda_i+\frac{1}{\alpha}(n+1-2i)\right].
\label{CS3}
\end{equation}

\section{$\textrm{WA}_{k-1}$ theories: definitions and main results}
\label{wtheory}
A complete construction of $W$ symmetry algebras  and 
their representation theories can be found  in \cite{Wtheory, Schoutens}.  Here we briefly review the main results of a particular family of $W$ theories, the $\textrm{WA}_{k-1}$ ones, already mentioned in the Introduction. A particular attention is given  to the series of minimal models $\textrm{WA}_{k-1}(p,p')$ with  $p$ and $p'$ coprime integers and  to the degeneration properties  of the operators  of the theory.

The  $\textrm{WA}_{k-1}$ model can be realized by a $(k-1)$-component Coulomb gas. The chiral currents $W^{(s)}$ can be expressed  in terms of  polynomials in derivatives of  a $k-1$ component free bosonic field $\vec{\varphi}(z)=(\varphi_{1},\varphi_{2},\cdots,\varphi_{k-1}),$ \cite{Wtheory} with the correlation functions normalized as:
\begin{eqnarray}
<\varphi_{a}(z,\bar{z})\varphi_{b}(z',\bar{z}')>=\log\frac{1}{|z-z'|^{2}}\delta_{ab}\nn\\
\end{eqnarray}
The  stress-energy operator $T(z)$ of the theory $\textrm{WA}_{k-1}(p,p')$ takes the form:
\begin{equation}
T(z)=-\frac{1}{2}:\partial\vec{\varphi}\partial\vec{\varphi}:+i\vec{\alpha}_{0}
\partial^{2}\vec{\varphi}
\label{stressenergy}
\end{equation}
The vector $\vec{\alpha}_{0}$ in the above equation is the background charge \cite{DotsenkoFateev} which is defined as:
\begin{equation}
\vec{\alpha_{0}}=\alpha_{0}\vec{\rho}=(\alpha_{+}+\alpha_{-})\sum_{a=1}^{k-1}\vec{\omega}_{a}
\end{equation}
where  the $\vec{\omega}_{a}$ are the fundamental weights of the $A_{k-1}$ Lie algebra and the parameters $\alpha_{\pm}$ are expressed in terms of $p$ and $p'$ by:
\begin{equation}
\alpha_{+}^2=\frac{p}{p'}\quad \alpha_{+}\alpha_{-}=-1;
\end{equation}
From the Eq.(\ref{stressenergy}), the central charge $c(p,p')$ of the $\textrm{WA}_{k-1}(p,p')$ models is:
\begin{equation}
 c(p,p') = k-1 - 12(\vec{\alpha}_{0})^2=(k-1)\left(1-\frac{k(k+1)(p-p')^2}{pp'}\right)
\label{centralcharge}
\end{equation}

The expressions for the other chiral currents $W^{(s)}$, $s=3,\cdots,k-1$ in terms of the derivatives of $\vec{\varphi}$ are more complicated, and they are not needed for our purposes.  What is important here is that the $W^{(s)}(z)$ ($s=3,\cdots, k-1$) classify, together with $T(z)$, all the operators  of the model in terms of primaries and descendants of the chiral algebra. The usual methods, combined with 
the available Coulomb gas representation, define the dimensions of primary operators  and their 
correlation functions. In this sense the conformal theories $\textrm{WA}_{k-1}$ are fully defined.

The  primary fields $\Phi_{\vec{\beta}}$ of the theory are  parametrized by the $k-1$  component vector $\vec{\beta}$. The behavior of a primary field $\Phi_{\vec{\beta}}$ under the action of the symmetry  generators $W^{(s)}$is encoded in the operator product expansions (OPE):
\begin{equation}
 T(z)\Phi_{\vec{\beta}}(w) =  \frac{\Delta_{\beta}
   \Phi_{\vec{\beta}}(w)}{(z-w)^2 } + \frac{\partial
   \Phi_{\vec{\beta}}(w)}{z-w} +\dots  \qquad \qquad   W^{(s)}(z)\Phi_{\vec{\beta}}(w)
 =  \frac{\omega^{(s)}_{\vec{\beta}} \Phi_{\vec{\beta}}(w)}{(z-w)^s } +\dots
\label{Wprimary} 
\end{equation}
The action of the chiral currents $T(z)$ and $W^{(s)}(z)$ can be expressed in terms of their modes $L_{n}$ and $W^{(s)}_{n}$ defined as:
\begin{equation}
T(z)\Phi_{\vec{\beta}}(w)=\sum_{n=-\infty}^{\infty} \frac{L_{n}\Phi_{\vec{\beta}}(w)}{(z-w)^{n+2}} \qquad \qquad \qquad W^{(s)}(z)\Phi_{\vec{\beta}}(w)=\sum_{n=-\infty}^{\infty} \frac{W_{n}^{(s)}\Phi_{\vec{\beta}}(w)}{(z-w)^{n+s}} 
\label{mode1}
\end{equation}
or equivantely:
\begin{equation}
L_{n}\Phi_{\vec{\beta}}(w)=\frac{1}{2\pi i}\oint_{\mathcal{C}_w}d\,z (z-w)^{n+1} T(z) \Phi_{\vec{\beta}}(w)  \qquad W_{n}^{(s)}\Phi_{\vec{\beta}}(w)=\frac{1}{2\pi i}\oint_{\mathcal{C}_w}d\, z (z-w)^{n+s-1}W^{(s)}(z) \Phi_{\vec{\beta}}(w)
\label{mode2}
\end{equation}
The conformal  dimension $\Delta_{\vec{\beta}}$ and the $\omega^{(s)}_{\vec{\beta}}$ are respectively  eigenvalues  of the zero modes $L_{0}$ and $W_{0}^{(s)}$ operators, $L_{0}\Phi_{\vec{\beta}}= \Delta_{\vec{\beta}} \Phi_{\vec{\beta}}$ and $W^{(s)}_{0}\Phi_{\vec{\beta}}= \omega^{(s)}_{\vec{\beta}}\Phi_{\vec{\beta}}$. The $\Delta_{\vec{\beta}}$ together with  the  set of the $k-2$ quantum numbers $\omega^{(s)}_{\vec{\beta}}$ characterize each  representation $\Phi_{\vec{\beta}}$. In particular the conformal dimension $\Delta_{\vec{\beta}}$ is given by:
\begin{equation}
\Delta_{\vec{\beta}}=\frac{1}{2}\vec{\beta}(\vec{\beta}-2\vec{\alpha}_{0})
\end{equation}
Notice also that, from the above definitions, $L_{-1}\Phi_{\vec{\beta}}(z)=\partial_{z} \Phi_{\vec{\beta}}(z)$.

The allowed values of the vectors $\vec{\beta}$ are defined by the
condition of complete degeneracy of the modules of $\Phi_{\vec{\beta}}(z)$ with
respect to the chiral algebra. 
 The Kac table is based on the weight lattice of the Lie algebra $A_{k-1}$ and the position of 
 the vectors $\vec{\beta}$ are found to be given by \cite{Wtheory}:
\begin{equation}
\vec{\beta}=\vec{\beta}_{(n_1,n_2\cdots n_{k-1}  \mid n_1',n_2'\cdots n_{k-1}')}=\sum_{a=1}^{k-1}\left((1-n_{a})
\alpha_{+} + (1-n_{a}')\alpha_{-}\right)\vec{\omega}_{a}
\end{equation}
Each primary operator $\Phi_{\vec{\beta}_{(n_1,n_2\cdots n_{k-1}  \mid n_1',n_2'\cdots n_{k-1})}}
\equiv \Phi_{(n_1,\cdots,n_{k-1}|n_1^{'},\cdots,n_{k-1}^{'})}$ is then characterized by the sets of integers $(n_1,\cdots,n_{k-1}|n_1^{'},\cdots,n_{k-1}^{'})$.
One can show that the representation $\Phi_{(n_1,\cdots,n_{k-1}|n_1^{'},\cdots,n_{k-1}^{'})}$ presents $k-1$ null vectors $\chi_{a}$ ($a=1,\cdots, k-1$) at level $n_{a}n_{a}'$. This directly generalizes the well known case of the degenerate representations of Virasoro algebra ($=\textrm{WA}_1$ algebra) \cite{diFrancesco}.

The principal domain of the Kac table contains the set of primary operators which form a closed fusion algebra, and is delimited as follows:
\begin{equation}
\sum_{a} n_a  \leq  p'-1 \qquad ;\qquad \sum_{a} n_a'  \leq  p-1 
\label{Kac}
\end{equation}
As it can be directly seen from the symmetries  of the conformal dimensions $\Delta_{\vec{\beta}}\equiv\Delta_{(n_1,\cdots, n_{k-1}|n_1'\cdots n^{'}_{k-1})}$, the operators in the Kac table are identified, up to a multiplicative factor \cite{TFTcf}, via the transformations  $\tau, \tau^2\cdots \tau^{k-1}$, $\Phi_{(n_1,\cdots,n_{k-1}|n_1^{'},\cdots,n_{k-1}^{'})}=\Phi_{\tau^{j}[(n_1,\cdots,n_{k-1}|n_1^{'},\cdots,n_{k-1}^{'})]}$ ($j=1,2,\cdots,k-1)$, where:
\begin{equation}
\tau[(n_1,\cdots,n_{k-1}|n_1^{'},\cdots,n_{k-1}^{'})]=(p'-\sum_{a=1}^{k-1} n_{a},n_1,\cdots,n_{k-2} | p-\sum_{a=1}^{k-1} n^{'}_{a},n_1^{'},\cdots,n_{k-2}^{'})
\label{identifications}
\end{equation}

\section{Parafermionic operators in $\textrm{WA}_{k-1}(k+1,k+r)$ theory, correlation functions and Jacks}
\label{paraop}

According to Coulomb gas rules, the  fusion of two operators $\Phi_{\vec{\beta}_1}$ and  $\Phi_{\vec{\beta}_2}$ produces an operator $\Phi_{\vec{\beta}_3}$ in the principal channel with $\vec{\beta}_3= \vec{\beta}_1+\vec{\beta}_2$,  namely  $\Phi_{\vec{\beta}_1}\times \Phi_{\vec{\beta}_2}=\Phi_{\vec{\beta}_3}+\cdots$ where the dots indicates the non-principal fusion channels. The non-principal channels follow the principal one by shifts realized by the roots $e_i$ ($i=1,\cdots,k-1$) of the $A_{k-1}$ Lie algebra.  A channel associated to an operator   which lies outside the Kac table (\ref{Kac})  does not enter in the fusion (i.e. the associated structure constant vanishes).  The operator algebra can then be easily  determined.

Let now consider the model $\textrm{WA}_{k-1}(k+1,k+r)$ where $p=k+1$ and $p'=k+r$. By using the Coulomb gas rules, one can verify that the set of operators:
\begin{equation}
\Psi_{i}= \Phi_{-\alpha_-\vec{\omega}_{i}} = \Phi_{-r\alpha_+\vec{\omega}_{k-i}}  \quad i=1,\dots,k-1, \quad \Delta_{i}=\frac{r}{2}\frac{ i(k-i)}{k}
\label{parafermionic_operators}
\end{equation}
form a subalgebra, namely $\Psi_{i}\times \Psi_{j}=\Psi_{i+j \;\mbox{mod}\;k}$. These fusion rules are only valid when $p=k+1$, because  in that case the fields $\Psi_{i}$ belong to the boundary of the Kac table, namely:
\begin{equation}
\Psi_{i} = \Phi_{\substack{(1,\dots,1 | 1,\dots ,1,2,1,\dots ,1)\\ \phantom{(1,\dots,1 | 1,\dots ,1,}\uparrow\phantom{,1,\dots ,1)} \\\phantom{(1,\dots,1 | 1,\dots ,1,}i\phantom{,1,\dots ,1)}}} = \Phi_{\substack{(1,\dots ,1,r+1,1,\dots ,1|1,\dots,1)\\ \phantom{(1,\dots ,1,}\uparrow\phantom{,1,\dots ,1|1,\dots,1)} \\ \phantom{(1,\dots ,1,}k-i\phantom{,1,\dots ,1|1,\dots,1)} }} \label{parafermionic_operators_index}
\end{equation}
and the usual fusion rules are truncated accordingly. The set of operators $\Psi_{i}$, which are degenerate representations of the $\textrm{WA}_{k-1}$ algebras, form then an associative $\mathbb{Z}_{k}^{(r)}$ parafermionic algebras \cite{FZ,JacobMathieu1,JacobMathieu2,ERS,Mathieu} with a fixed  central charge given by $c(k+1,k+r)$, see Eq.(\ref{centralcharge}). In particular the $\Psi_{i}$ operators can be identified with the parafermionic chiral currents with $\mathbb{Z}_k$ charge equal to $i$. Notice that the dimensions of the fields $\Psi_{i}$ and $\Psi_{k-i}$ are the same. This reflects the fact that the correlation function of $\Psi_{i}$ operators are symmetric under the conjugation of charge $i\to k-i$. 

In the following  we will use quite often the notation $\Psi$ for the field $\Psi_{1}$ or its conjugate $\Psi_{k-1}$, and we will use $\Delta$ and $\omega^{(3)}$ for the corresponding eigenvalues of $L_0$ and $W^{(3)}_0$. The correlation function $\langle  \Psi(z_1) \dots \Psi(z_{n}) \rangle$ we will consider denotes then the correlation function   $\langle  \Psi_1(z_1) \dots\Psi_1(z_{n}) \rangle= \langle  \Psi_{k-1}(z_1) \dots \Psi_{k-1}(z_{n}) \rangle$. It should be noted that for these correlators to be non-zero, $n$ should be a multiple of $k$.

It has been conjectured   \cite{ FJMM,FJMM2, BernevigHaldane2,BernevigW} that these $n-$point correlation functions $\langle  \Psi(z_1) \hdots
\Psi(z_{n}) \rangle$ can be written in terms of a single Jack polynomial. Namely the  conjecture is that:
 \begin{eqnarray}
 \langle  \Psi(z_1) \hdots  \Psi(z_{n}) \rangle=P_n^{(k,r)} (\{z_i\}) \prod_{i<j}\left(z_i-z_j \right)^{-r/k}.
\label{CFT_FQH}
\end{eqnarray}
The  $P_n^{(k,r)}$ is the follwing Jack in $n$ variables:  
\begin{equation}
\quad P_{n}^{(k,r)} (\{z_i\}) = J_{\lambda}^{-(k+1)/(r-1)} (\{z_i\}) 
\label{conj1}
\end{equation}
where:
\begin{equation} 
\lambda=[\underbrace{N_{\phi},\dots , N_{\phi}}_{\text  k \textrm{ times}}, \underbrace{N_{\phi}-r,\dots,N_{\phi}-r}_{\text  k \textrm{ times}},\cdots ,\underbrace{r,\dots ,r}_{\text  k \textrm{ times}}]
\label{conj2}
\end{equation}
and:
\begin{equation}
N_{\phi}=\frac{r(n-k)}{k}
\label{conj3}
\end{equation}
Notice that the above notations has been inherited from the  FQH notations where  the $N_{\phi}$ denotes  the magnetic flux. The $P_{n}^{(k,r)}$ describes lowest Landau level bosonic particles at filling fraction $\nu=r/k$.

\section{Second order differential equations for the $n-$point functions  $\langle\Psi(z_1) \Psi(z_2)\cdots \Psi(z_n)\rangle$ }
We consider the $n$-point correlation function $\langle\Psi(z_1) \Psi(z_2)\cdots \Psi(z_n)\rangle$  of the $\textrm{WA}_{k-1}(k+1,k+r)$
theory and we show that these functions satisfy a particular second order differential
equation. We can prove then that these correlation functions are written in terms of a single Jack.

\subsection{$\textrm{WA}_{k-1}$ symmetry: Ward identities}
The possible form of a general  correlation function is restricted by the $\textrm{WA}_{k-1}$ symmetry. More specifically, each correlation function satisfies a  Ward identity associated to each symmetry current $T(z)$ and $W^{(s)}$, $s=3,\cdots,k$.  These identities  can be easily obtained from the Eq.(\ref{Wprimary}).   For  the stress energy tensor $T(z)$ and $W^{(3)}(z)$ we have:
\begin{eqnarray}
\langle T(z) \Psi(z_1)\cdots \Psi(z_n) \rangle &=& \sum_{j=1}^{n} \left(\frac{\Delta}{(z-z_j)^2} \langle \Psi(z_1)\cdots \Psi(z_n) \rangle+ \frac{1}{(z-z_j)}\langle \Psi(z_1)\cdots \partial_{j}\Psi(z_j)\cdots \rangle \right)  \label{Ward_vir}\\
\langle W^{(3)}(z) \Psi(z_1)\cdots \Psi(z_n) \rangle &=& \sum_{j=1}^{n} \left( \frac{\omega^{(3)}}{(z-z_j)^3} \langle \Psi(z_1)\cdots \Psi(z_n)\rangle +  \frac{1}{(z-z_j)^2}\langle \Psi(z_1)\cdots W_{-1}^{(3)}\Psi(z_j)\cdots \rangle  \right.\nonumber \\
&&+ \left. \frac{1}{(z-z_j)}\langle \Psi(z_1)\cdots W^{(3)}_{-2}\Psi(z_j)\cdots \Psi (z_n)\rangle \right)
\label{Ward_W}
\end{eqnarray}
By demanding that the functions $\langle T(z) \Psi(z_1)\cdots \Psi(z_n) \rangle$ and $\langle W^{(3)}(z) \Psi(z_1)\cdots \Psi(z_n) \rangle$ are regular at $z\to \infty$ and using the transformations law of the $T(z)$ and $W^{(3)}(z)$ under a conformal map, one can easily verify that the functions $\langle T(z)\dots\rangle$ and $\langle W^{(3)}(z)\dots\rangle$ behaves respectively like:
\begin{equation}
T(z)\sim \frac{1}{z^4} \quad \textrm{ and } \quad W^{(3)}(z) \sim \frac{1}{z^6} \quad \textrm{as } z\to \infty
\label{asymp}
\end{equation}
Comparing the asymptotics (\ref{asymp}) and the Ward identities (\ref{Ward_vir})-(\ref{Ward_W}), one can derive a set of relations satisfied by the correlation functions $\langle \Psi(z_1)\cdots \Psi(z_n)\rangle$. For instance, using the Eq.(\ref{asymp}) in Eq.(\ref{Ward_vir}) one has:
\begin{eqnarray}
\sum_{j=1}^{n}  \partial_{j} \langle \Psi(z_1)\cdots \Psi(z_n) \rangle &=& 0 \label{L^-}\\
\sum_{j=1}^{n}  \left(z_j \partial_{j} + \Delta \right) \langle \Psi(z_1)\cdots \Psi(z_n) \rangle  &=& 0 \label{L^z}\\
\sum_{j=1}^{n}  \left(z_j^2   \partial_{j}+ 2 z_j \Delta_{j}\right) \langle \Psi(z_1)\cdots \Psi(z_n) \rangle &=& 0 \label{L^+}
\end{eqnarray}
The relations \eqref{L^-}-\eqref{L^+} take the form of simple differential equations and impose the invariance of the correlation function $ \langle \Psi(z_1)\cdots \Psi(z_n) \rangle$ under global conformal transformations \cite{diFrancesco}. As shown in \cite{TFTcf}, anagously to the case of the conformal symmetry, one can derive a set of relations associated to the symmetry generated by the spin $3$ current $W^{(3)}(z)$. Again, using Eq.(\ref{asymp}) in Eq.(\ref{Ward_W}) one obtains the following 5 relations:
\begin{eqnarray}
\sum_{j=1}^{n}   \langle \Psi(z_1)\cdots W_{-2}^{(3)}\Psi(z_j) \cdots \Psi(z_n) \rangle &=& 0 
\label{w1}\\
\sum_{j=1}^{n}   \langle \Psi(z_1) \cdots \left(z_j W_{-2}^{(3)} + W_{-1}^{(3)} \right)\Psi (z_j)\cdots \Psi(z_n) \rangle &=& 0 \label{w2}\\
\sum_{j=1}^{n}  \langle \Psi(z_1) \cdots \left(z^2_j W_{-2}^{(3)} +2 z_j W_{-1}^{(3)}+\omega^{(3)} \right)\Psi (z_j)\cdots \Psi(z_n) \rangle &=& 0 \label{w3}\\
\sum_{j=1}^{n}  \langle \Psi(z_1) \cdots \left(z^3_j W_{-2}^{(3)} +3 z^2_j W_{-1}^{(3)}+3 z_j \omega^{(3)} \right)\Psi (z_j)\cdots \Psi(z_n) \rangle &=& 0 \label{w4}\\
 \sum_{j=1}^{n}    \langle \Psi(z_1) \cdots \left(z^4_j W_{-2}^{(3)} +4 z^3_j W_{-1}^{(3)}+6 z^2_j \omega^{(3)} \right)\Psi (z_j)\cdots \Psi(z_n) \rangle &=& 0\label{w5}
\end{eqnarray}

We stress that the above set of relations are very general constraints of the $\textrm{WA}_{k-1}$ theory. Although we have written down these relations for the specific case of the correlation function under consideration, any primary fields  correlation function of the $\textrm{WA}_{k-1}$ theory satisfies  constraints of the same kind.

\subsection{$\textrm{WA}_{1}(3,2+r)$: minimal models of Virasoro algebra}
\label{vir_minmodel}
The $\textrm{WA}_{1}(3,2+r)$ theories, corresponding to $k=2$, coincide with the minimal model $M(3,2+r)$.
Notwithstanding the fact that in this case the relation between
 correlation functions in Virasoro minimal models and Jacks is quite well known,  we discuss
 briefly the $k=2$  case since we shall investigate the more complicated cases $k>2$
 in an analogous fashion.

As  it has been observed in \cite{JacobMathieu1,JacobMathieu2},  the
$\mathbb{Z}_2$ parafermionic operator $\Psi_1$ (\ref{parafermionic_operators}) coincides with  the
$\Phi_{(1\mid 2)}$ operator , $\Psi=\Psi_1=\Phi_{(1\mid 2)}$. The operator $\Psi$ has
conformal dimension $\Delta=\Delta_{(1\mid 2)}=r/4$ and  the $\Psi \Psi$  fusion
realizes the $\mathbb{Z}_2^{(r)}$ parafermionic algebra with central charge
$c=1-2(r-1)^2/(2+r)$, see Eq.(\ref{centralcharge}). Moreover, the operator $\Psi$ satisfies  a second level null vector $\chi_{2}$ condition \cite{diFrancesco}:
\begin{equation}
\left(L_{-2}-\frac{3}{r+2}L_{-1}^2\right) \Psi=0
\label{vir_nullvector}
\end{equation}
The degeneracy condition (\ref{vir_nullvector})  implies that correlation functions containing  $\Psi$ obey a second order differential equation. 
Let us consider a general correlation function $\langle\Phi(z)\Phi_1(w_1)\Phi_2(w_2)\cdots\rangle $  involving some  primary operators
$\Phi,\Phi_1, \cdots$. By using Eq.(\ref{Wprimary}), Eq.(\ref{mode2}) and the Cauchy theorem, one can
always express the action of the $L_{n}$ modes for $n \leq 1$ on the  primary $\Phi$ in
terms of differential operators acting on the others primaries $\Phi_i$:
\begin{eqnarray}
\langle
(L_{n}\Phi(z))\Phi_{1}(w_1)\Phi_{2}(w_2)\cdots\rangle&=&-\sum_{j}(n+1)(w_j-z)^{n}\Delta_{j}\langle
\Phi(z)\Phi_{1}(w_1)\Phi_{2}(w_2)\cdots\rangle \nonumber
\\&& -\sum_{j}(w_j-z)^{n+1}\partial_{w_{j}}\langle
\Phi(z)\Phi_{1}(w_1)\Phi_{2}(w_2)\cdots\rangle
\label{vir_diff}
\end{eqnarray}
Notice that the above Eq.(\ref{vir_diff}) is a more general form of the Virasoro Ward identity (\ref{Ward_vir}).

Suming over the singular vector equation (\ref{vir_nullvector}) resulting from each field $\Psi$ and using the Eq.(\ref{vir_diff}),
 one obtains the following differential equation satisfied by the  $n-$point correlation function $\langle
\Psi(z_1) \Psi(z_2)\cdots \Psi(z_n)\rangle$. Defining $\mathcal{H}^{\mbox{vir}}$ as:
\begin{equation}
\mathcal{H}^{\mbox{vir}}(r)\hat{=}\sum_{i=1}^{n}\left(z_i^2\partial_{i}^2-\frac{r+2}{3}\sum_{j\neq i}\left(\frac{r}{4}\frac{z_i^2}{(z_i-z_j)^{2}}+\frac{z_i^2 \partial_{j}}{z_i-z_j}\right)\right)
\label{H_vir}
\end{equation}
one has:
\begin{equation}
\mathcal{H}^{\mbox{vir}}(r)\langle \Psi(z_1) \Psi(z_2)\cdots \Psi(z_n)\rangle=0.
\end{equation}
Using the result from the Appendix, this can be put in the following form:
\begin{equation}
\mathcal{H}^{\textrm{WA}_{1}}(r)\langle \Psi(z_1) \Psi(z_2)\cdots \Psi(z_n)\rangle=0.
\end{equation}
where
\begin{eqnarray}
& &\mathcal{H}^{\textrm{WA}_{1}}=\sum_i  \left( z_i \partial_i \right)^2 +\gamma_1(r) \sum_{i\neq j} \frac{z_j^2}{(z_j-z_i)^2} + \gamma_2(r)\sum_{i\neq j}\frac{z_iz_j(\partial_j-\partial_i)}{(z_j-z_i)} +n\gamma_3(r)\nn \\
& & \gamma_1=-\frac{r(r+2)}{12} \qquad \gamma_2=\frac{r+2}{6}\qquad \gamma_3=-\frac{r(r-1)}{12} 
\label{diff_eq_wa1}
\end{eqnarray}
Let us introduce the function $\phi^{(r,k)}(\{z_i\})$:
\begin{equation}
\phi^{(r,k)}(\{z_i\})\hat{=}\prod_{i<j}(z_i-z_j)^{r/k} 
\label{vand}
 \end{equation}
After some algebraic manipulations, conjugation with the function $\phi^{(r,2)}$ transforms the second-order differential
operators $\mathcal{H}^{\textrm{WA}_{1}}$, defined in Eq.(\ref{diff_eq_wa1}), into the Calogero Hamiltonian
$\mathcal{H}^{\mbox{CS}}(\alpha)$ 
Eq.(\ref{CS}):
\begin{equation}
[\phi^{(r,2)}(\{z_i\})]\mathcal{H}^{\textrm{WA}_{1}}[\phi^{(r,2)}(\{z_i\})]^{-1}=\mathcal{H}^{\mbox{CS}}(\alpha)
-E(r)
\label{conj_k2a}
\end{equation}
with:
\begin{equation}
\alpha=-\frac{3}{r-1}\quad E(r)=\frac{1}{36}r n (n-2)\left[2+n+r(2n -5)\right] 
\label{conj_k2b}
\end{equation} 
One can easily verify by comparing  Eqs.(\ref{conj_k2a})-(\ref{conj_k2b}) with Eqs(\ref{CS})-(\ref{CS3}) that  
 the Eqs.(\ref{conj1})-(\ref{conj3}) are verified for $k=2$. It is important to stress that the Jack solution (\ref{CS2}) of the eigenvalue equation (\ref{CS}) is the only solution with  monodromies consistent with the OPE of the operators $\Psi$, i.e. with the $\mathbb{Z}_{2}^{(r)}$ parafermionic algebra.

In the more general case of the $\textrm{WA}_{k-1}$ theories with $k=3,4\cdots$, it is
in general impossible to  write down differential equations for correlation
functions containing  one
completely degenerate fields and other arbitrary fields. Generally speaking,
 the null-vector conditions of the  $\textrm{WA}_{k-1}$ theory
present for $k>2$, in addition to the $L_{n}$ Virasoro modes, the modes
$W_{n}^{(s)}$ of the higher spin currents. The action of
the modes $W_{n}^{(s)}$ do not have a geometrical interpretation, i.e. they can not be written as differential operators.
This is the essential difficulty in the analysis of the $W$ theory correlation functions.

In the following, we will closely  use the approach  of \cite{TFTcf}. We will first give a detailed analysis of the case $k=3$. From the degeneracy properties of the parafermionic operators $\Psi$, we show that the correlation functions involving $n$ operators $\Psi$ satisfy a second order differential equation. This equation allows to prove the conjecture \eqref{CFT_FQH} for the theory $\textrm{WA}_{2}(4,3+r)$ (i.e. $k=3$). Then we show how to generalize this result for the general case.

\subsection{$\textrm{WA}_{2}(4,3+r)$ models}

The chiral algebra contains the stress energy operator $T(z)$ and the
$W^{(3)}(z)$ current of spin $3$. The explicit form of the $\textrm{WA}_{2}$ algebra,
written in terms of the commutators between the chiral current modes, is:
\begin{eqnarray}
\left[L_n,L_m \right] & = & (n-m)L_{n+m} + \frac{c}{12}n(n^2-1)\delta_{n+m,0} \\
\left[L_n,W^{(3)}_m \right] & = & \left( 2n-m\right)W^{(3)}_{n+m} \\
\left[W^{(3)}_n,W^{(3)}_m \right] & = & \frac{16}{22+5c} (n-m) \Lambda_{n+m}  + \frac{c}{360} n(n^2-1)(n^2-4) \delta_{n+m,0}  \nonumber \\ 
& & + (n-m)\left[\frac{1}{15}(n+m+2)(n+m+3)-\frac{1}{6}(n+2)(m+2) \right] L_{n+m} 
\label{W3algebra}
\end{eqnarray}
with
\begin{eqnarray}
 \Lambda_n &= &d_n L_n + \sum_{m=-\infty}^{\infty} :L_m L_{n-m}: \\
d_{2m} & = & \frac{(1-m^2)}{5} \\
d_{2m-1} & = & \frac{(1+m)(2-m)}{5}
\end{eqnarray}
The $A_{2}$ weight lattice is two-dimensional and the representations of the
$\textrm{WA}_2$ algebra $\Phi_{\vec{\beta}}=\Phi_{(n_1,n_2\mid n_1',n_2')}$ are indexed by the couple of integers
$(n_1,n_2\mid n_1',n_2')$. The Kac table is delimited by:
\begin{equation}
 n_1+n_2 \leq p'-1 \qquad n_1'+n_2' \leq p-1 \label{Kac_WA2}
\end{equation}

The $\Psi_{1}$ and the
$\Psi_{2}$ operators, which  are  identified  in the Eq.(\ref{parafermionic_operators_index}) as:
\begin{eqnarray}
\Psi_{1} & = & \Phi_{(1,r+1\mid 1,1)} = \Phi_{(1,1\mid 2,1 )} \\
\Psi_{2} & = &\Phi_{(r+1,1\mid 1,1)} = \Phi_{(1,1\mid 1,2 )}, \\
\label{Psi_iden}
\end{eqnarray}
generate the $\mathbb{Z}_{3}^{(r)}$ parafermionic theory. In the Eq.(\ref{Psi_iden}) the identifications (\ref{identifications}) are used.
The operators $\Psi_{1}$ and $\Psi_{2}$ are $\mathbb{Z}_3$-charge conjugates and have the same dimension $\Delta$:
\begin{equation}
\Delta=\Delta_{(1,1\mid 2,1)}=\Delta_{(1,1\mid 1,2)} =\frac{r}{3}
\label{dim_k3}
\end{equation}
Notice that the $\Psi_{1}$ and $\Psi_{2}$ are distinct $\textrm{WA}_2$ representations as one can directly see from the fact that the  associated  $W_{0}^{(3)}$ eigenvalues   $\omega^{(3)}_{(1,1\mid
  1,2)}$  and $\omega^{(3)}_{(1,1\mid 2,1)}$,  see Eq.(\ref{Wprimary}),   have opposite sign,  $\omega^{(3)}_{(1,1\mid
  1,2)}=-\omega^{(3)}_{(1,1\mid 2,1)}$ \cite{TFTcf}. Their value is given by:
  \begin{equation}
 \left(\omega^{(3)}\right)^2 = \frac{2\Delta^2}{9}\left(\frac{32}{22+5c}(\Delta+\frac{1}{5}) -\frac{1}{5} \right) 
  \label{omegavaluek3}
  \end{equation}

\subsubsection{$\textrm{WA}_2$ null-vectors conditions}
The fields $\Psi_{1}$ and $\Psi_{2}$, identified in Eq.(\ref{Psi_iden}) respectively to the degenerate representations $\Phi_{(1,1\mid 2,1 )}$ and 
$\Phi_{(1,1\mid 1,2 )}$, are expected to have two null-vectors  at level $1$ and $2$. From the commutation relations (\ref{W3algebra}), one can show \cite{Wtheory,TFTcf} that the fields $\Psi_{1}$ and $\Psi_{2}$, defined in Eq.(\ref{Psi_iden}), satisfy the following null-vector conditions:
\begin{eqnarray}
\left(W^{(3)}_{-1} -\frac{3\omega^{(3)}}{2\Delta}L_{-1} \right)\Psi &=& 0 \nonumber \\
\left( W^{(3)}_{-2} -\frac{12\omega^{(3)}}{\Delta(5\Delta+1)}L_{-1}^2 - \frac{6\omega^{(3)}(\Delta+1)}{\Delta(5\Delta+1)}L_{-2} \right)\Psi  &=& 0
\label{W_3nullvector}
\end{eqnarray}
where $\omega^{(3)}$ stands for $\omega^{(3)}_{(1,1\mid 2,1)}$ (respectively $\omega^{(3)}_{(1,1\mid 1,2)}$) when $\Psi_{1}$ ($\Psi_{2}$) is concerned.
We remark that the fields $\Psi_{1}$ ($\Psi_{2}$) satisfy  an additional third level null-vector conditions which directly comes from the conditions (\ref{W_3nullvector}) and the algebra (\ref{W3algebra})\cite{TFTcf}. For our purposes we do not need such condition.

Here we are interested in the $n$-point correlation function $\langle \Psi(z_1) \cdots \Psi(z_n)\rangle$, see Section(\ref{paraop}). As it is explicitly shown in Eq.(\ref{W_3nullvector}), the modes of the additional current $W^{(3)}(z)$ appear in the null-vector conditions (\ref{W_3nullvector}). 

 \subsubsection{Second order differential equation for $\langle \Psi(z_1)\cdots \Psi(z_n) \rangle$}
\label{proof_a2}

We show here  that the null-vector conditions (\ref{W_3nullvector}) allow us to derive a second-order differential equation for  $\langle \Psi(z_1) \cdots \Psi(z_n)\rangle$. To take care of the modes $W^{(3)}_{-2}$ and $W^{(3)}_{-1}$, one can use any of the relations (\ref{w1})-(\ref{w5}), together with the null-vector conditions (\ref{W_3nullvector}), to obtain a relation involving purely the Virasoro modes $L_{-2}$ and $L_{-1}$($=\partial$). This allow to obtain five different differential equations for the $n-$point functions. As suggested by the results known for the Jacks \cite{Bernevig}, all these differential equations are not independant and can be obtained form one another by commutation with \eqref{L^-}-\eqref{L^+}. Of particular interest to us is the following equation, obtained by using Eqs.(\ref{W_3nullvector}) in Eq.(\ref{w3}):
\begin{eqnarray}
0 = \sum_{j=1}^{n} \langle \Psi(z_1) \cdots \left(z_j^2 W_{-2}^{(3)} + 2z_j W_{-1}^{(3)}+ \omega^{(3)} \right)\Psi(z_j) \cdots \Psi(z_n)\rangle = & & \\
 \sum_{j=1}^{n} \langle \Psi(z_1)\cdots \left[\frac{-8 a}{5\Delta+1}z_j^2\left( \partial_{j}^2 - \frac{\Delta+1}{2}L_{-2}\right)- 2 a z_j \partial_j - \frac{2}{3} a \Delta \right]\Psi(z_j)\cdots \Psi(z_n)\rangle &  & 
\end{eqnarray}
where $a=-3 \omega^{(3)}/(2\Delta)$.  Notice that the constant $a$ factorizes in the above equations, and we are left with:
\begin{equation}
\sum_{j=1}^n \langle \Psi(z_1) \left[z_j^2\left( \partial_j^2 - \frac{\Delta+1}{2}L_{-2} \right)  + \frac{5\Delta+1}{4} z_j \partial_j  + \frac{\Delta(5\Delta+1)}{12}   \right]\Psi(z_j)  \dots \Psi(z_n)\rangle = 0 
\end{equation}
This means that the sign of $ \omega^{(3)}$ does not modify the differential equation. This is consistent with the fact that, as previously mentioned, correlation functions are invariant under the charge conjugation $\Psi_{1}\leftrightarrow\Psi_{2}$.
Taking into account the following relations:
\begin{eqnarray}
 \sum_k z_j\partial_j \langle \Psi(z_1) \dots \Psi(z_n)\rangle &= &-n\Delta\langle \Psi(z_1) \dots \Psi(z_n)\rangle \\
\langle \Psi(z_1)\dots L_{-2}\Psi(z_j) \dots \Psi(z_n)\rangle & = &\sum_{\substack{i=1 \\i \neq j}}^n \left(\frac{\Delta}{(z_j-z_i)^2} + \frac{\partial_i}{z_j-z_i} \right) \langle \Psi(z_1)   \dots \Psi(z_j)  \dots \Psi(z_n)\rangle,
\end{eqnarray}
and using the Eq.(\ref{dim_k3}),we can write down the second-order differential equation for $\langle \Psi(z_1) \dots \Psi(z_n)\rangle$ where the coefficients $\gamma_i(r)$ ($i=1,2,3$)  are  given as functions of $r$.   We have found:
\begin{equation}
\mathcal{H}^{\textrm{WA}_2}\langle \Psi(z_1) \dots \Psi(z_n)\rangle = 0
\label{diff_eq_a2}
\end{equation}
where $\mathcal{H}^{\textrm{WA}_2}$ is:
\begin{eqnarray}
\mathcal{H}^{\textrm{WA}_2}&=&\sum_{j=1}^{n}  \left( z_j   \partial_j \right)^2   +\gamma_1(r) \sum_{i\neq j} \frac{z_j^2}{(z_j-z_i)^2}+ \gamma_2\sum_{i\neq j}\frac{z_iz_j(\partial_j-\partial_i)}{(z_j-z_i)} + n \gamma_3(r)\\
 \gamma_1(r)&=& -\frac{r(r+3)}{18} \quad \gamma_2(r)= \frac{3+r}{12} \quad \gamma_3=- \frac{r(4 r-3)}{27}
\label{ham_k3}
\end{eqnarray}

Analogously to what we have seen in Sec.(\ref{vir_minmodel}), we use the function $\phi^{(r,3)}(\{z_i\})$ defined in Eq.(\ref{vand}) to transform the above second-order differential equation into the Calogero Hamiltonian (\ref{CS})(see Appendix):
\begin{equation}
[\phi^{(r,3)}(\{z_i\})]\mathcal{H}^{\textrm{WA}_2}(r)[\phi^{(r,3)}(\{z_i\})]^{-1}=\mathcal{H}^{\mbox{CS}}(\alpha)
-E(r)
\label{conj_k3a}
\end{equation}
with:
\begin{equation}
\alpha=-\frac{4}{r-1}\quad E(r)=\frac{n r}{216} (-3+n)(9-21 r + n (3 + 5 r))
\label{conj_k3b}
\end{equation} 
By comparing  Eqs.(\ref{conj_k3a})-(\ref{conj_k3b}) with Eqs(\ref{CS})-(\ref{CS3}), it is straightforward to see  that  Eqs.(\ref{conj1})-(\ref{conj3}) are verified for $k=3$. As we have said for the case $k=2$,  the Jack solution (\ref{CS2}) of the eigenvalue equation (\ref{CS}) is the only solution consistent with the single-channel fusion rules of the $\mathbb{Z}_{3}^{(r)}$ parafermionic algebra, i.e. it is the only polynomial solution.

\subsection{$\textrm{WA}_{k-1}(k+1,k+r)$ models}
We complete the proof of the Eqs.(\ref{conj1})-(\ref{conj3}) for general $k$, i.e. for the general theory $\textrm{WA}_{k-1}(k+1,k+r)$.
The parafermions operators $\Psi_{1}$ and $\Psi_{k-1}$ are identified with the following primary fields:
\begin{eqnarray}
\Psi_{1}&=&\Psi_{(1,1\cdots,r+1 \mid 1, \cdots,1)}= \Phi_{(1,1,\dots,1|2,1,\dots1)}\\
\Psi_{k-1}&=&\Psi_{(r+1,1\cdots,1 \mid 1, \cdots,1)}= \Phi_{(1,1,\dots,1|1,1,\dots2)}
\end{eqnarray}
with conformal dimension $\Delta$:
\begin{equation}
\Delta=\Delta_{(1,1,\dots1|2,1,\dots1)}=\Delta_{(1,1,\dots, 1|1,1,\dots2)}=\frac{r}{2}\frac{k-1}{k}
\label{dim_k}
\end{equation}
where we have used the identifications (\ref{identifications}). In the following we set $\Psi \hat{=} \Psi_{1}$ and we compute the $n-$point function 
$\langle \Psi(z_1)\cdots \Psi(z_n) \rangle$. The results we obtain are valid also for the $n-$point correlation functions of the conjugate field $\Psi_{k-1}$.

The field $\Psi$ is expected to have $k-2$ null-vectors at level $1$ and one null-vector at level $2$. But the situation is slighlty more complex since the descendants of these null states also decouple from the theory, and in general the embedding of these null-state modules is non trivial. Nevertheless, using the characters of the $\textrm{WA}_{k-1}$ theories \cite{Wtheory,Schoutens}, or equivalently the reflections along the roots in the Coulomb gas language, it is rather straightforward to count the number of remaining independent fields at a given level. In particular we showed that the representation module of $\Psi_{1}$ (or $\Psi_{k-1}$) only has one state at level one, and two independent states  at level two.  This statement does not hold for the other parafermionic fields $\Psi_n$, $n=2,\dots, k-2$: in that case there are three independents states at level two. This is not surprising because the conjecture relating parafermionic correlation functions and Jack polymomials only holds for the parafermions with the lowest dimension: $\Psi_1$ and $\Psi_{k-1}$.

For these two fields, the first two levels are completely spanned by the Virasoro modes, and all the additional modes corresponding to the currents $W^{(s)}$, $s=3,\cdots k-1$ only appear in null-vectors. In particular the field $W_{-2}^{(3)}\Psi$ and $W_{-1}^{(3)}\Psi$ can be written as linear combination of Virasoro modes:
\begin{eqnarray}
\left(W^{(3)}_{-1} + a L_{-1} \right)\Psi = 0 \nonumber \\
\left( W^{(3)}_{-2} + \mu L_{-1}^2 +\nu L_{-2} \right)\Psi = 0
\label{null_vector_k}
\end{eqnarray}
where the constants $ a$, $\mu$ and $\nu$ are computed below. This result is consistent with the works \cite{Hornfeck,Bajnok}, where it was shown that starting precisely from  the null-vector conditions (\ref{null_vector_k}) (and a chain of other conditions  for the other currents) as hypotheses, one can rebuild the $\textrm{WA}_{k-1}$ algebra.

The constants $a$, $\mu$ and $\nu$ can be determined by acting with positive Virasoro modes on the null vectors (\ref{null_vector_k}). We have obtained:  \begin{eqnarray}
a & = & -\frac{3\omega^{(3)}}{2\Delta}\\
\mu & = & a\frac{2(2\Delta+c)}{(-10\Delta+16\Delta^2+2c\Delta+c)} = a \frac{2k(1+k)}{(rk^2+k^2-2k-4r)}  \\
\nu & = & a \frac{16\Delta(\Delta-1)}{(-10\Delta+16\Delta^2+2c\Delta+c)} = -\mu \frac{2(k+r)}{k(1+k)}
\label{coeff_null}
\end{eqnarray}
where $\omega^{(3)}=\pm \omega^{(3)}_{(1,1,\cdots1\mid 2,1,\cdots,1)}$  for $\Psi=\Psi_{\pm1}$.

Replacing  $k=3$ in the above equation, one obtains the ones given in the Eq.(\ref{W_3nullvector}). Notice however that the coefficients given above are different from the ones  obtained by replacing the values of $\Delta$ of the Eq.(\ref{dim_k})  in the coefficients  of the Eq.(\ref{W_3nullvector}). This is quite natural as one expects  that the presence  of the higher spin currents in the chiral algebra modifies the coefficients of the null-vector conditions. 

The differential equation satisfied for $\langle \Psi(z_1)\cdots \Psi(z_n)\rangle$ for the general theory $\textrm{WA}_{k-1}$ can then be obtained in the same fashion as in the case $k=3$, see Section (\ref{proof_a2}). By  using Eqs.(\ref{null_vector_k})-(\ref{coeff_null}) into Eq.(\ref{w3}) we obtain:
\begin{equation}
\mathcal{H}^{\textrm{WA}_{k-1}}\langle \Psi(z_1) \dots \Psi(z_n) \rangle = 0
\end{equation}
where the differential operator $\mathcal{H}^{\textrm{WA}_{k-1}}$, whose coefficients are given as functions of $r$ and $k$,  is defined as:
\begin{eqnarray}
& &\mathcal{H}^{\textrm{WA}_{k-1}}=\sum_i  \left( z_i \partial_i \right)^2 +\gamma_1(k,r) \sum_{i\neq j} \frac{z_j^2}{(z_j-z_i)^2} + \gamma_2(k,r)\sum_{i\neq j}\frac{z_iz_j(\partial_j-\partial_i)}{(z_j-z_i)} +n\gamma_3(k,r)\\
& & \gamma_1=-\frac{r(rk-r+k^2-k)}{k^2(k+1)} \qquad \gamma_2=\frac{r+k}{k(k+1)}\qquad \gamma_3=-\frac{r(k-1)(2rk-k-2r)}{6k^2} 
\label{diff_eq_wak}
\end{eqnarray}
As we have seen in the case $k=3$, see Section (\ref{proof_a2}) the constant $a$ can is simplified during the derivation of the above equation. The Eq.(\ref{diff_eq_wak}) is then independent of the sign of $\omega^{(3)}=\pm \omega^{(3)}_{(1,1,\cdots1\mid 2,1,\cdots,1)}$. Once 
again, this is consistent with the invariance of the parafermionic correlation functions under  charge conjugation ($i\to k-i$). As expected, we recover the pure Virasoro case when $k=2$.

Using the function $\phi^{(r,k)}$, defined in Eq.(\ref{vand}), we can transform the above differential equation into the Calogero Hamiltonian.  We have:
\begin{equation}
[\phi^{(r,k)}(\{z_i\})]\mathcal{H}^{\textrm{WA}_{k-1}}[\phi^{(r,k)}(\{z_i\})]^{-1}=\mathcal{H}^{\mbox{CS}}(\alpha)
-E(r)
\label{conj_ka}
\end{equation}
with:
\begin{equation}
\alpha=-\frac{k+1}{r-1}\quad E(r)=\frac{n r (k - n) [-2 n r + k^2 (-1 + 2 r) - k (n - r + n r)]}{6 k^2 (1 + 
   k)}
\label{conj_kb}
\end{equation} 
By comparing  Eqs.(\ref{conj_ka})-(\ref{conj_kb}) with Eqs(\ref{CS})-(\ref{CS3}), it is straightforward to see  that  Eqs.(\ref{conj1})-(\ref{conj3}) are verified for each $k$. This completes the proof of the conjecture relating Jack wavefunctions to $\textrm{WA}_{k-1}(k+1,k+r)$ theories.

\section{Conclusion}
In this paper we computed the $n-$point correlation function of the field $\Psi_{1}=\Phi_{(1,\cdots,1\mid 2,1\cdots,1)}$ and of the field  $\Psi_{k-1}=\Phi_{(1,\cdots,1\mid 1,1\cdots,2)}$ belonging to the Kac table of the minimal model $\textrm{WA}_{k-1}(k+1,k+r)$. By using the Ward identities associated to the spin $3$ curent $W^{(3)}(z)$ and the degeneracy properties of the $\Psi_{1}$ and $\Psi_{k-1}$ representations, we showed that their $n-$point  correlation functions  satisfy a second order differential equation. This equation can be transformed into a Calogero Hamiltonian with negative rational coupling $\alpha=-(k+1)/(r-1)$. This completes the proof of the conjecture which states that the $n-$point correlation functions of $\Psi_{1}$ ($\Psi_{k-1}$) can be written in term of a single Jack polynomial.

{\it Acknowledgements}: The authors thanks  E.~Ardonne, Vl.~Dotsenko and N.~Regnault for very helpful discussions. B.E. aslo wishes to thank B.A.~Bernevig for explaining the nature of the additional differential equations satisfied by the Jack polynomials. R.S. acknowledges conversations with N.~Cooper, Th.~Jolicoeur, V.~Fateev and  S.~Ribault. 

\section{Appendix}

In order to derive the Hamilontians $\mathcal{H}^{\textrm{WA}_{k-1}}$ from the null vector conditions, the following relation is quite useful:
\begin{equation}
\sum_{i \neq j} \left( \frac{z_i^2\partial_j}{z_i-z_j} \right)  \langle \Psi(z_1) \dots \Psi(z_n)\rangle =  \left( n\Delta -  \frac{1}{2}\sum_{i\neq j}\frac{z_iz_j(\partial_j-\partial_i)}{(z_j-z_i)}\right) \langle \Psi(z_1) \dots \Psi(z_n)\rangle \label{useful_relation}
\end{equation}
In order to derive this relation it is convenient to introduce the following differential operators:
\begin{eqnarray}
\mathcal{D}&  =&  \sum_{i=1}^n z_i \partial_i \\
\mathcal{T} & =& \sum_{i=1}^n \partial_i \\
\mathcal{O} & =& \sum_{\substack{i,j=1\\i\neq j}}^n\frac{z_iz_j(\partial_i-\partial_j)}{(z_i-z_j)} 
\end{eqnarray}
one has: 
\begin{eqnarray}
\sum_{\substack{i,j=1\\i\neq j}}^n \left( \frac{z_j^2\partial_i}{z_j-z_i} \right) & = &\sum_{i\neq j}  \left(z_j \partial_i + \frac{z_i z_j \partial_i}{z_j - z_i}  \right) \\
& = & \sum_j z_j \sum_{i \neq j}\partial_i - \frac{1}{2}\sum_{i\neq j}\frac{z_iz_j(\partial_i-\partial_j)}{(z_i-z_j)} \\
& = &  \sum_j z_j \left[-\partial_j + \sum_i \partial_i \right] - \frac{1}{2}\mathcal{O} \\
& = &  -\mathcal{D} +\left( \sum_j z_j\right) \mathcal{T}  -\frac{1}{2}\mathcal{O}
\end{eqnarray}
The action on a correlation function $\langle \Psi(z_1) \dots \Psi(z_n)\rangle$ greatly simplifies since:
\begin{eqnarray}
 \mathcal{T}\langle \Psi(z_1) \dots \Psi(z_n)\rangle & = & 0 \\
\mathcal{D} \langle \Psi(z_1) \dots \Psi(z_n)\rangle & = & -n \Delta \langle \Psi(z_1) \dots \Psi(z_n)\rangle
\end{eqnarray}
and one gets \eqref{useful_relation}.


\begin{thebibliography}{10}

\bibitem{diFrancesco}
P.~di Francesco,  P.~Mathieu and D.~Senechal, 
\newblock {\it Conformal Field Theory}, Springer NewYork (1997).

\bibitem{Dotsi_cours}
Vl.~Dotsenko
\newblock {\it Series de cours sur la theorie conforme} http://cel.archives-ouvertes.fr/cel-00092929/en/

\bibitem{cardy1}
J.~Cardy
\newblock Phys.Lett. B {\bf 582} 121(2004) 

\bibitem{cardy2}
B.~Doyon and J.~Cardy
\newblock J. Phys. A 40 2509 (2007) 


\bibitem{Calogero}
F.~Calogero
\newblock J.Math. Phys {\bf 10}, 2191 (1969);

\bibitem{Sutherland71}
B.~Sutherland
\newblock J.Math. Phys {\bf 12}, 246 (1971);{\bf 12}, 251 (1971)

\bibitem{Sutherland72}
B. Sutherland, 
\newblock  Phys. Rev. A4, 2019 (1971); 5, 1372 (1972).







\bibitem{FJMM}
B.~ Feigin, M.~Jimbo, T. ~Miwa and E.~Mukhin,
\newblock International Mathematics Research Notices 1223 (2002).

\bibitem{FJMM2} 
B.~Feigin, M.~Jimbo, T.~Miwa and E.~Mukhin, 
\newblock International Mathematics Research Notices 1015 (2003); arXiv:math/0209042



\bibitem{BernevigHaldane1}
B.~A.~ Bernevig and F.~D.~M.~ Haldane, 
\newblock Phys. Rev. Lett. {\bf 100}, 246802 (2008).

\bibitem{BernevigHaldane2}
B.A. Bernevig and F.D.M. Haldane, 
\newblock Phys. Rev. Lett. {\bf 101}, 246806 (2008).

\bibitem{Wtheory}
V. A. Fateev and S. L. Lykyanov, 
\newblock  Int. J. Mod. Phys. A{\bf 3} 507 (1988).


\bibitem{Schoutens}
P.~Bouwknegt and K.~Schoutens
\newblock Phys.Rept. 223 (1993) 183-276

\bibitem{TFTcf}
V. A. Fateev and S. L. Litvinov, 
\newblock  JHEP0711:002 (2007).

\bibitem{TFTcf2}
V. A. Fateev and S. L. Litvinov, 
\newblock  JHEP0901:033 (2009).
\bibitem{MooreRead}
G.~Moore and N.~Read,
\newblock Nucl. Phys. {\bf B 360}, 362 (1991).

\bibitem{ReadRezayi}
N.~Read and E.~Rezayi,
\newblock  Phys. Rev. {\bf B 59}, 8084 (1999).

\bibitem{BernevigW}
B. ~A.~Bernevig, V.~Gurarie, S.~H.~ Simon
\newblock  J. Phys. A: Math. Theor. {\bf 42} 245206 (2009) 

\bibitem{Ardonne}
E. Ardonne, 
\newblock  Phys. Rev. Lett. {\bf 102}, 180401 (2009).


\bibitem{Macdonald}
I.~G.~Macdonald, 
\newblock {\it Symmetric functions and Hall polynomials} , 2nd ed., Oxford University
Press, New York, 1995.


\bibitem{DotsenkoFateev} 
Vl. S.~Dotsenko and V.A.~Fateev,\newblock 
                         Nucl.~Phys.~{\bf B324}~312~(1984),
                         ~{\bf B251}~691~(1985);
                        \newblock ~Phys. Lett. ~{\bf B154}~291~(1985).





\bibitem{FZ}
A.~ Zamolodchikov and V.~ Fateev,
\newblock Sov..Phys. JETP {\bf 62}, 215-225 (1985).

\bibitem{JacobMathieu1}
P. Jacob and P. Mathieu,
\newblock Nucl.Phys. B733 205-232 (2006)


\bibitem{JacobMathieu2}
P. Jacob and P. Mathieu
\newblock   Physics letters B. {\bf 627}, 224 (2005).

\bibitem{ERS}
B.~Estienne, N.~Regnault and R.~Santachiara
\newblock Nucl. Phys. B824, 539-562(2010)

\bibitem{Mathieu}
P.~Mathieu
\newblock J.Phys.A:Math.Theor.42 375212 (2009)





\bibitem{Bernevig} 
B.A. Bernevig, private communication



\bibitem{Hornfeck}
K.~Hornfeck
\newblock Nucl.~Phys.~B {\bf 411} 307 (1994) 

\bibitem{Bajnok}
Z.~Bajnok
\newblock Lett.Math.Phys. {\bf 49},325 (1999)













\bibitem{zomo32}
V~. A.~ Fateev and A.~ B.~ Zamolodchikov,
\newblock  Theot. Math. Phys. {\bf 71} 451 (1987).
 
\bibitem{zamo_infinite} 
 A.~ B.~ Zamolodchikov, 
\newblock   Theor. Math. Phys. {\bf 63} 1205 (1985).
 


\end{thebibliography}
\end{document}